\documentclass[manuscript]{aastex631}
\usepackage{graphicx,epsfig,fancyhdr,epsf,txfonts,epstopdf}% Include figure
\graphicspath{{./}{figures/}}
\shorttitle{Plasma Emission Induced By Electron Beam}
\shortauthors{Chen et al.}
\begin{document}

\title{Plasma Emission Induced By Electron Beam in Weakly Magnetized Plasmas}

\correspondingauthor{Yao Chen}
\email{yaochen@sdu.edu.cn}

\author{Yao Chen}
\affiliation{Institute of Frontier and Interdisciplinary Science, Shandong University, Qingdao, Shandong, 266237, People's Republic of China.}
\affiliation{Institute of Space Sciences, Shandong University, Shandong, 264209, People's Republic of China.}

\author{Zilong Zhang}
\affiliation{Institute of Space Sciences, Shandong University, Shandong, 264209, People's Republic of China.}
\affiliation{Institute of Frontier and Interdisciplinary Science, Shandong University, Qingdao, Shandong, 266237, People's Republic of China.}

\author{Sulan Ni}
\affiliation{Institute of Space Sciences, Shandong University, Shandong, 264209, People's Republic of China.}
\affiliation{Institute of Frontier and Interdisciplinary Science, Shandong University, Qingdao, Shandong, 266237, People's Republic of China.}

\author{Chuanyang Li}
\affiliation{Institute of Frontier and Interdisciplinary Science, Shandong University, Qingdao, Shandong, 266237, People's Republic of China.}
\affiliation{Institute of Space Sciences, Shandong University, Shandong, 264209, People's Republic of China.}

\author{Hao Ning}
\affiliation{Institute of Space Sciences, Shandong University, Shandong, 264209, People's Republic of China.}
\affiliation{Institute of Frontier and Interdisciplinary Science, Shandong University, Qingdao, Shandong, 266237, People's Republic of China.}

\author{Xiangliang Kong}
\affiliation{Institute of Space Sciences, Shandong University, Shandong, 264209, People's Republic of China.}
\affiliation{Institute of Frontier and Interdisciplinary Science, Shandong University, Qingdao, Shandong, 266237, People's Republic of China.}

\begin{abstract}
Previous studies on the beam-driven plasma emission process were done mainly for unmagnetized plasmas. Here we present fully-kinetic electromagnetic particle-in-cell simulations to investigate such process in weakly-magnetized plasmas of the solar corona conditions. The primary mode excited is the beam-Langmuir (BL) mode via the classical bump-on-tail instability. Other modes include the whistler (W) mode excited by the electron cyclotron resonance instability, the generalized Langmuir (GL) waves that include a superluminal Z-mode component with smaller wave number $k$ and a thermal Langmuir component with larger $k$, and the fundamental (F) and harmonic (H) branches of plasma emission. Further simulations of different mass and temperature ratios of electrons and protons indicate that the GL mode and the two escaping modes (F and H) correlate positively with the BL mode in intensity, supporting that they are excited through nonlinear wave-wave coupling processes involving the BL mode. We suggest that the dominant process is the decay of the primary BL mode. This is consistent with the standard theory of plasma emission. Yet, the other possibility of the Z+W$\rightarrow$O--F coalescing process for the F emission cannot be ruled out completely.
\end{abstract}


\section{Introduction} \label{sec:intro}

The standard plasma emission (PE) mechanism, first proposed more than 6 decades ago
\citep{1958SvA.....2..653G}, is a multi-stage nonlinear process including:
(1) efficient excitation of Langmuir (L) turbulence by electron beams
through the kinetic bump-on-tail instability; (2)
scattering of Langmuir waves by ion-acoustic (IA) wave or ion
density inhomogeneities to generate the fundamental (F) O-mode
radiation and/or the backward-propagating Langmuir waves, noted as
L $\pm$ IA$\rightarrow$O--F and L $\pm$ IA$\rightarrow$L$'$; and (3)
resonant coupling of forward- and backward- propagating
Langmuir turbulence to generate the harmonic (H) radiation, noted as
L+L$'$$\rightarrow$H. Extensive studies on PE have been carried out with the original
theoretical framework being largely maintained \citep[e.g.,][]{melrose_emission_1980,Melrose1987,cairns_fundamental_1987,1994ApJ...422..870R, 2012JGRA..117.4106S, 2013JGRA..118.4748L, 2014SoPh..289..951L, 2014AGUFMSM13E4215C, 2014cosp...40E.446C,2015A&A...584A..83T,2017PNAS..114.1502C,2019JGRA..124.1475H}.

Despite significant progresses, it is very challenging to verify the complete PE mechanism
with PIC simulation, partially due to limits of computational resources
since such simulation should adopt large-enough domain and long-enough duration
so as to support relevant wave modes. The frequency of the fundamental emission of the O
mode is at or very close to its cutoff around the plasma oscillation frequency ($\sim \omega_\mathrm{pe}$). This corresponds to a small wave number and a long wavelength. The IA mode, determined
by the ion dynamics, has a relatively low frequency (and a long period). In addition, the simulation
is necessarily multi-dimensional to consider the wave coupling process, and number
of macro-particles per cell should be as large as possible so as to reduce the noise
level. With these constraints, earlier studies present inconclusive
or even contradictory statements regarding whether and which of the proposed
PE processes does occur in their simulations. \citet{2015A&A...584A..83T} presented a
critical review on these earlier studies, pointing out that
some studies did not conduct the convergence test and/or disentangle
contributions from non-escaping electromagnetic modes, thus
relevant PE process has not been conclusively verified.

\citet{2015A&A...584A..83T} concluded that simulations with a dense
beam ($n_\mathrm{b}/n_\mathrm{0}$ $\sim$ 0.05, the density ratio of the beam to
background electrons) prohibit the L+IA$\rightarrow$O--F process
while those with a dilute beam ($n_\mathrm{b}/n_\mathrm{0}$ $\sim$ 0.0057) allow
such process. This is in line with the dispersion analysis of the beam-plasma system \citep{1989PhFlB...1..204C} that indicates mismatch of the coalescence conditions in the
presence of a dense beam. This point has been used to explain why some earlier
studies fail to verify the complete PE process.

\citet{2015A&A...584A..83T} demonstrated the excitations of
the F and H emissions from a single-beam plasma system via
fully-kinetic PIC simulations, in accordance with the standard
PE theory. Yet, it is questionable regarding whether the simulation time is long
enough and the domain is large enough to simulate the
low-frequency IA and the long-wavelength O--F emission. Their study does not
take the magnetization effect into account. This is equivalent to
assume that the plasma emission process is unaffected by
magnetization effect, and waves such as the whistler (W) mode and
the so-called electromagnetic superluminal Z mode are irrelevant.

\citet{2019JGRA..124.1475H} however found the F emission in
their simulation is hardly discernible though they did adopt a dilute
beam with spatial domain (simulation time) larger (longer) than that of
\citet{2015A&A...584A..83T}, within similar unmagnetized beam-plasma system.

On the other hand, latest theoretical studies extended earlier analysis on the kinematics of electrostatic Langmuir decay from unmagnetized to magnetized plasmas \citep{2013PhRvL.110r5001L,2018PhPl...25h2309C}, and found that such decay is always kinematically permitted and can proceed for very fast streams and generate Langmuir waves with very small wave numbers that are the Z-mode component of the so-called generalized Langmuir mode (GL) in magnetized plasmas \citep[see, e.g.,][]{1986islp.book.....M,2000PhPl....7.3167W}. This removes the kinematic constraints from earlier unmagnetized theory and may affect the evolution of the beam-plasma system and relevant PE process. The predicted decay process is yet to be verified with PIC simulations.

According to recent PIC simulations of PE for weakly-magnetized plasmas
energized by energetic electrons with the loss-cone type distributions \citep{2020ApJ...891L..25N, 2021ApJ...909L...5L}, the
fundamental O--mode (O--F) emission can be generated through
the coalescence of the almost-counter propagating
Z and W modes that are excited through the electron cyclotron
resonance instability (ECMI).
Follow-up studies verified the occurrence of such Z+W$\rightarrow$O--F process
through the wave-pumping method of PIC simulation \citep{2021PhPl...28d0701N}.
It is intriguing to figure out whether the W and Z modes and their coalescing process can
occur in a magnetized beam-plasma system.

Thus, it is demanding to do PIC simulations of the beam-plasma interaction in magnetized plasmas, to verify the occurrence of the above Langmuir decay process and to clarify the role of electromagnetic modes such as W and Z in generating PEs. This is the main purpose of the present study.

\section{Simulation Method and Parameter Setup}

The beam-plasma system was simulated with the open-source
Vector PIC (VPIC) code on supercomputers operated by the Beijing Super-Cloud
Computing Center (BSCC). It is a fully kinetic electromagnetic and relativistic
code released by the Los Alamos National Labs, run in two spatial
dimensions ($x, z$) with three velocity components using periodic boundary conditions \citep{2008PhPl...15e5703B,bowers20080,2009JPhCS.180a2055B}.
The background magnetic field is set to be $\vec B_0$ (=$B_0 \hat e_z$), and
the wavevector $\vec k$ is in the $xOz$ plane. The plasmas consist of three
components, including background electrons and protons with
the Maxwellian distribution, and the electron beam with the
following velocity distribution function (VDF, see Figure~\ref{fig1}c)
\begin{equation}
f_\mathrm{e} = A_\mathrm{e} \exp(-\frac{u^2_\perp}{2u^2_{0}} - \frac{(u_\parallel - u_\mathrm{d})^2}{2u^2_{0}})
\end{equation}
where $u_\parallel$ and $u_\perp$ are the parallel and perpendicular components
of the momentum per mass, $u_0$ is the thermal velocity of energetic electrons,
and $A_\mathrm{e}$ is the normalization factor. The density ratio of beam-background electrons
is set to be 0.01. All particles distribute homogeneously in
space initially. The drift speed of the electron beam is set to be $u_\mathrm{d}=0.2824c$ (~20 keV), and the ratio of $\omega_\mathrm{pe}/\Omega_\mathrm{ce}$ to be 10. These
parameters are consistent with the general conditions
of the solar corona.

The domain of the simulation is taken to be $L_x = L_z =2048\ \Delta$, where
the grid spacing $\Delta  = 3.25\ \lambda_\mathrm{de}$, $\lambda_\mathrm{de}$ is the Debye
length of the background electrons. The simulation lasts for 2000 $\omega_\mathrm{pe}^{-1}$.
The resolvable range of $|k|$ is $[0.52,535]$ $\Omega_\mathrm{ce}/c$, and the range of $\omega$
is $[0.04,32]$ $\Omega_\mathrm{ce}$ (for the time interval of $1500$ $\omega_\mathrm{pe}^{-1})$.
The number of macro-particles per cell is taken to be 2000 for
the background electrons and 1000 for both protons
and the beam. The zero-current condition is maintained initially.

\section{Results and Analysis}
We first present details of wave excitation for the reference case (Case R) with realistic proton-electron mass ratio ($m_\mathrm{p}/m_\mathrm{e}=1836$ and equal temperature ($T_\mathrm{p} = T_\mathrm{e} = 2$ MK); then we compare these results with other cases of different $m_\mathrm{p}/m_\mathrm{e}$ and $T_\mathrm{p}/T_\mathrm{e}$ to shed lights on the role of ion-related perturbations and physical connection among various wave modes.

\subsection{Wave analysis for Case R}
\begin{figure*}[ht]
 \centering
 \includegraphics[width=0.99\linewidth]{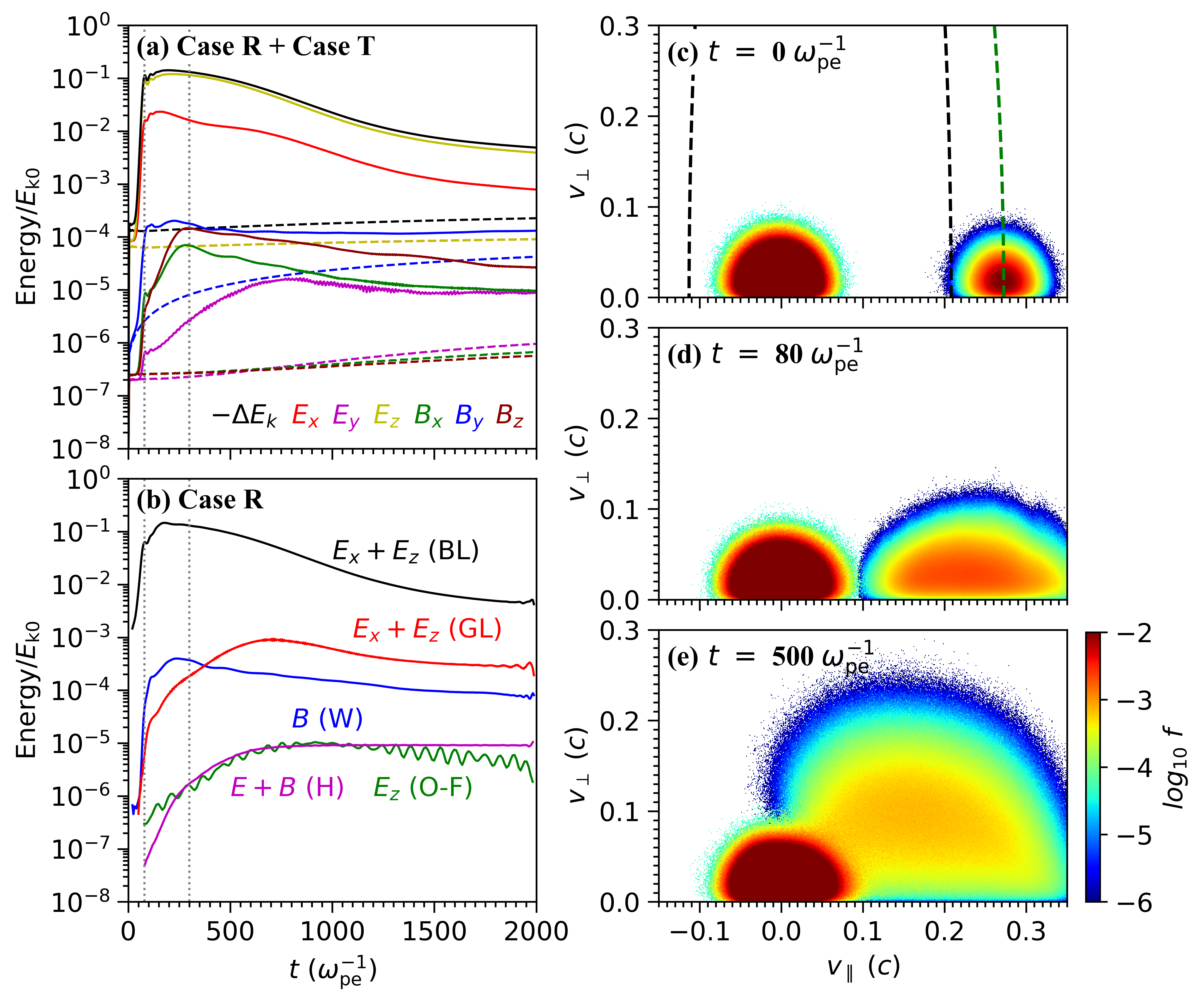}
  \caption{(a) energy variations of the six field components and the negative change of the
total electron energy ($-\Delta E_k$) for \textbf{Case R  (solid lines) and Case T (dashed lines)}; (b) energy variations of specific field
components of various wave modes for Case R (BL for the beam-Langmuir mode,
W for whistler, O--F for the fundamental and H for the harmonic PE); panels (c--e)
illustrate the VDFs at three moments. Overplotted in panel (c) are two resonance curves
of the W mode with the following ($\omega, k, \theta$) values: [$0.1\ \Omega_\mathrm{ce}$, $4.5\ \Omega_\mathrm{ce}/c$,  $30^\circ$] and [$0.3\ \Omega_\mathrm{ce}$, $12.3\ \Omega_\mathrm{ce}/c$,  $60^\circ$]. The two dashed lines in panels (a--b) represent $t\ =\ 80\ \omega_\mathrm{pe}^{-1}$ and $t\ =\ 300\ \omega_\mathrm{pe}^{-1}$.}\label{fig1}
\end{figure*}

We first present the energy profiles of various field
components for Case R, in panel a of Figure~\ref{fig1}, overplotted by
those for the corresponding thermal case (Case T), to show the significance of wave growth in
the unstable beam-plasma system. All field components of Case R
are stronger in energy than those of the thermal case by at
least 2--4 orders in magnitude, further analysis on energy profiles of
individual wave modes (panel b) verifies their significant
enhancements over the corresponding thermal noise signals.
This is important since in PIC simulations numerical noises organize themselves
as signals along dispersion curves that might be mistaken as wave excitation \citep[c.f.,][]{2015A&A...584A..83T}.

According to the energy profiles, we split the whole simulation
into three stages (see Figure~\ref{fig1}a--c): 0--80 $\omega_\mathrm{pe}^{-1}$ for Stage I characterized by
the rapid rise of $E_x$ and $E_z$ in energy, corresponding to the
growth of the primary BL mode;
80--300 $\omega_\mathrm{pe}^{-1}$ for Stage II characterized
by the gradual rise of $B_x$ and $B_z$ associated with the excitation
of the whistler (W) mode; and 300--1500 $\omega_\mathrm{pe}^{-1}$ for Stage III corresponding
to the saturation and gradual damping of various wave modes. At the
end of Stage I the value of $-\Delta E_k$ is about 12\% of $E_{k0}$, which
increases slightly to the maximum at $\sim$ 150 $\omega_\mathrm{pe}^{-1}$, then
declines gradually as a result of the return of wave energy to electrons.
In panel b of Figure~\ref{fig1}, we plot the energy variation of specific field component
of different wave modes, as will be discussed later.

Panels c--e present VDF maps of electrons for
Case R at $t =$ 0, 80, and 500 $\omega_\mathrm{pe}^{-1}$.
During 0--80 $\omega_\mathrm{pe}^{-1}$ (Stage I),
the beam electrons are decelerated causing the rapid rise of $E_x$ and $E_z$,
and diffuse toward larger $v_\perp$ later. Note that similar behavior of VDF
diffusion (towards larger $v_\perp$) has been indicated by \citet{2020PhPl...27b0702H} and
\citet{2021SoPh..296...42M} in terms of 3-D generalization of the beam-plasma interaction. They concluded
that electrons can diffuse significantly in angle and thus form broad distribution similar
to that illustrated here (rather than forming a plateau of VDF as implied by the previous 1-D picture).
The VDF does not change significantly after $\sim$ 500 $\omega_\mathrm{pe}^{-1}$.

\begin{figure*}[ht]
 \centering
 \includegraphics[width=0.99\linewidth]{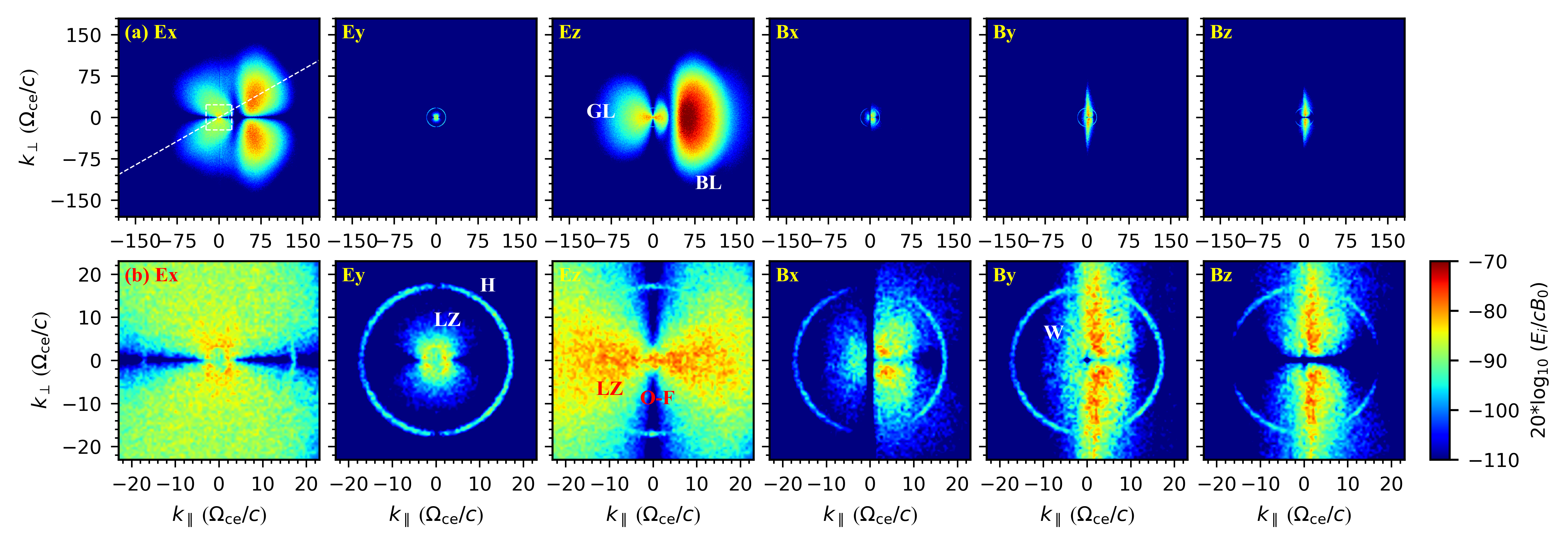}
  \caption{Intensity maps of the six field components in the wave vector $\vec k$ space
  for Case R. Right columns of panels are the zoom-in version of the squared region plotted in the top-left panel. The dashed line indicates the propagation angle of ($\theta = 30^\circ$) along which
  the $\omega, k$ dispersion maps are plotted in Figure~\ref{fig3}. }\label{fig2}
\end{figure*}

In Figure~\ref{fig2}, we present the wave-energy maps in the
wave vector ($\vec k$) space of the six electric ($\vec E$)
and magnetic ($\vec B$) field components for Case R. The maps show the
intensity maxima of waves at corresponding $\vec k$. The nature
of the modes can be easily identified from the analytic dispersion curves
plotted in Figure~\ref{fig3} and the accompanying movie. The strongest feature is carried by $E_z$ (panel c),
corresponding to the primary forward-propagating electrostatic BL mode that is
excited via the well-known bump-on-tail instability.
The BL mode extends $\sim \pm 60^\circ$ away from the parallel direction
with significant perpendicular component $E_x$. In panels a and c, the two
secondary wave enhancements distribute along one smooth dispersion curve,
which represent the so-called generalized Langmuir (GL) mode with two
components: the backward-propagating thermal Langmuir (LT) waves with
larger $k$ and the Z-mode (LZ) component with smaller $k$. The angular patterns of the
two components (LT and LZ) are similar to that of the primary BL wave.

Significant enhancements on the $\vec B$ dispersion map are mainly associated with
the W mode. The nice circular pattern represents the harmonic (H) radiation.
In the middle part of the $E_y$ dispersion, the enhanced features with low $k$
correspond to the mixture of the electromagnetic LZ-mode and the fundamental
O mode (O--F) emission. It is not possible to separate them with this type of dispersion.

In Figure~\ref{fig3}, we present the $\omega$--$k$ dispersion analysis
along propagation angles ($\theta$) with strong wave enhancement. As mentioned, this
figure should be combined with Figure~\ref{fig2} to tell the mode properties. The left column of panels (a) shows the strongest BL mode, within the $\omega$--$k$ range of [9.0, 10.5] $\Omega_{ce}$ and [40, 150] $\Omega_{ce}/c$.
As seen from panels a and b, the GL mode exists in the range of [9.6, 10] $\Omega_{ce}$ and [-20, 20] $\Omega_{ce}/c$ for its Z-mode component
and [10, 10.4] $\Omega_{ce}$ and [-90, -20] $\Omega_{ce}/c$ for its thermal component, and the O--F mode
appears in the range of [-5, 5] $\Omega_{ce}$ and [10, 10.2] $\Omega_{ce}/c$.

The O--F mode is mainly carried by the parallel electric field $E_z$
while the LZ mode by $E_x$ and $E_z$. Both modes present
signatures in $E_y$ due to their
electromagnetic nature, and both have phase speeds larger than $c$. Note that the over-plotted dashed lines
are dispersion curves given by the classical cold plasma magnetoionic
theory. The good correlation indicates that the thermal effect is not
important to the two superluminal modes.

As seen from the third row of panels (c), the W mode is also strongly excited.
To demonstrate its excitation mechanism, we plot
two sets of resonance curves of the ECMI process onto Figure~\ref{fig1}c. Both curves pass through positive gradient region of the beam distribution, this supports that it is
excited via ECMI. Note that the exact growth rates
depend on an integral along the resonance ellipse with an
integrand that depends on the partial derivatives of the VDF
with respect to both parallel and perpendicular velocity.
The W mode is dominated by the three magnetic-field
components. The energy profile of this mode (Figure~\ref{fig1}b)
rises rapidly during the early stage of the beam-plasma interaction, similar
to that of the BL mode, this also supports it is excited directly.
Note that according to \citet{1999PhPl....6.2862S} and \citet{2017PhPl...24g2116A},
in plasmas with a large $\omega_\mathrm{pe}/\Omega_\mathrm{ce}$
the W mode can be directly excited through the ECMI process with $n=1$ where $n$ represents
the harmonic number of ECMI, rather than the Landau resonance
with $n=0$. This is consistent with our analysis.

The fourth row of panels (d) present the H emission, corresponding
to the circular patten in Figure~\ref{fig2}. Its ($\omega$ and $k$) ranges
are [19.8, 20.2] $\Omega_\mathrm{ce}$ and [17, 18] $\Omega_\mathrm{ce}/c$.
According Figure~\ref{fig1}b, the total field energy of the H emission is about the same
as that of the $E_z$ energy of the O--F emission.

We examine the temporal development of various wave modes with energy curves for specific field
component(s) (see Figure~\ref{fig1}b), which are calculated by integrating field energy
within a specified spectral range along the corresponding dispersion curve according to the Parseval's theorem. The BL mode presents the strongest and fastest growth during the first stage, the W mode gets excited at a slower pace mainly in the second stage, and the GL mode grows gradually with time. Both PE modes (O--F and H) grow gradually before reaching the saturation level, similar to the energy profile of the GL mode.

We also checked the resonance curves of the GL mode and the two PE modes (not shown).
Only a minor part of the GL mode has resonance curves passing
through VDF region of positive gradient, meaning that this mode
cannot be excited directly through ECMI. For the O--F and H emissions,
the resonance curves either do not exist (no solutions to the matching equation)
or could not pass through any region of positive gradient of VDF.
Therefore, neither the two PE modes can be excited via ECMI.
This agrees with the standard mechanism that PEs in plasmas
with $\omega_\mathrm{pe}/\Omega_\mathrm{ce} \gg 1$ involve nonlinear
process of three-wave coupling process.

To further explore the excitation mechanism of GL and the two PE
modes, we present dispersion relation of the IA mode (see panel (e) of
Figure~\ref{fig3}), there exist bidirectional features of wave enhancements
along the standard dispersion curve of IA. Note that due to the
limited simulation time and spatial domain, the IA mode can at most
be marginally resolved with Case R. This hinders us from further analysis.

\subsection{Wave analysis for cases with different $m_\mathrm{p}/m_\mathrm{e}$ and $T_\mathrm{p}/T_\mathrm{e}$}
Significant excitations of LZ and PE (O--F and H) modes is the most important result
of the study. To dig out the underlying physics, we conducted additional numerical
experiments with $m_\mathrm{p}/m_\mathrm{e} (=1836/9, 1836 \times 25, 1836 \times 100)$ and $T_\mathrm{p}/T_\mathrm{e} (=1)$ and another case with $T_\mathrm{p}/T_\mathrm{e} =0.2$ and $m_\mathrm{p}/m_\mathrm{e} =1836$.
The solutions are presented in Figures~\ref{fig4} and \ref{fig5}. In Figure~\ref{fig4} we show the
Fourier spectra in the wave vector space for
density fluctuations of electrons ($\delta n_\mathrm{e}$)
and protons ($\delta n_\mathrm{p}$). The bottom panels
are for the low-frequency regime of the ($\omega, k$) spectra.
In this regime, the two species have identical spectra so only those
for protons are presented. Figure~\ref{fig5} (and the accompanying movie) presents the $\omega-k$ spectra along selected propagating
direction to illustrate the intensities of the BL mode, the GL mode including its LZ and LT
components, and the O--F mode. The W mode spectra do not change much with the two
ratios, thus not shown.

As seen from Figure~\ref{fig4}, the main difference between the two sets of spectra is the presence
of high-frequency electrostatic BL mode in the $\delta n_\mathrm{e}$ spectra. The other features
belong to low-frequency fluctuations associated with protons with charge neutrality being well maintained. In Figure~\ref{fig4}, the major feature of the $\vec k$-space spectra with
$m_\mathrm{p}/m_\mathrm{e} = 1836$ and $1836/9$ is the vertical wave enhancement which is
the quasi-perpendicularly propagating IA mode (coupled with the
cyclotron motion). These features are separated
from the BL mode by about 30 $\Omega_\mathrm{ce}/c$ in $k_\parallel$, therefore
they could not participate the three-wave coupling process involving the BL mode to generate
the LZ and the O--F modes whose $k$ values are in general less than 10 $\Omega_\mathrm{ce}/c$.

There exist two additional weaker features of IA along parallel (anti-parallel) to
oblique propagating direction, which (and the above quasi-perpendicular mode) get invisible in
cases with $m_\mathrm{p}/m_\mathrm{e} > 1836 \times 25$ since their frequency is too low to be resolved.
Note that with $T_\mathrm{p}/T_\mathrm{e} = 1$ the IA may be
dissipated by the Landau damping, thus cannot grow to stronger levels.
To verify this, we conducted another simulation with $T_\mathrm{p}/T_\mathrm{e} = 0.2$ and
$m_\mathrm{p}/m_\mathrm{e} = 1836$.

Comparing Figures~\ref{fig4} and \ref{fig5}, we made two observations: (1)
The three modes (BL, GL, and O--F) get stronger with increasing $m_\mathrm{p}/m_\mathrm{e}$ if
$m_\mathrm{p}/m_\mathrm{e} \le 1836 \times 25$, in other words, the two modes (GL including its
LZ and LT components and O--F) correlate positively in intensity with BL.
This reveals underlying physical connection between BL and GL/O--F;
if $m_\mathrm{p}/m_\mathrm{e}$ is further increased to $1836 \times 100$ then these modes do not change anymore, while the IA features disappear completely. (2) Comparing Case R with the $T_\mathrm{p}/T_\mathrm{e} = 0.2$ case reveals that the IA mode gets stronger for smaller $T_\mathrm{p}/T_\mathrm{e}$ (due to weaker damping effect), yet the BL and GL modes remain at levels close to those of Case R while the O--F mode manifests weak enhancement along quasi-parallel to oblique propagating direction.

\section{Conclusions and Discussion}
This study presents fully-kinetic electromagnetic PIC simulations of the interaction
of a beam of energetic electrons with weakly-magnetized plasmas that are
characterized by solar coronal conditions. The main purpose is to
investigate wave excitation and plasma emission mechanism. Waves
that are directly excited include the forward-propagating beam-Langmuir (BL)
mode via the bump-on-tail instability and the whistler (W) mode via
the cyclotron resonance instability. Waves that are excited through
nonlinear wave-wave interactions include the generalized Langmuir (GL)
mode that consists of the superlunimal Z-mode component at small
wave number and the backward-propagating thermal component
at large wave number. In addition, significant excitation of
the O--mode fundamental (O--F) and harmonic (H) plasma emissions are also observed.

%By comparing cases with different proton-electron mass and temperature ratios,
%we deduce that the dominant process accounting for the excitation
%of the GL wave and the F emission is the decay of the primary BL
%wave,

%and it is the nonlinear resonant coupling between the
%backward-propagating thermal component of the Langmuir wave and the BL wave to generate
%the H emission, consistent with the standard theory of plasma emission.
%Yet, the possibility of the coalescence of the Langmuir Z-mode wave
%with the W mode wave to generate the O--F emission cannot be ruled out.

Our simulations support that the d (decay) process (BL - IA)
of the primary BL mode plays a key role in generating both components of the GL mode
(the LZ component at smaller $k$ and the LT component at larger $k$) and the O--F emission, while the u (coalescence) process (BL + IA) may be not important here.
This argument is based on the following three aspects:
\textbf{(1) the 3 modes (BL, GL, and IA) are observed in simulations
for reasonable ion masses and ion temperatures,} (2) the two modes (GL and O--F)
correlate positively with the BL wave in intensity,
and (3) they maintain a strong level even in the absence of IA, and do not
get enhanced significantly when the IA mode is enhanced.

% In addition, the H emission is
%likely generated via the nonlinear resonant coupling between the
%backward-propagating thermal component of the Langmuir wave and the BL wave,
%consistent with the standard theory of plasma emission.
%, and (3) the two modes
%(GL and O--F) cannot be excited directly via the ECMI process according to
%the resonance analysis and further calculations of their linear growth rates.

The present study provides the first demonstration of the
generation of the Langmuir-Z mode in terms of the decay of the
primary beam-driven Langmuir wave in weakly-magnetized plasmas, using fully-electromagnetic PIC simulation. This agrees with the latest
theoretical extension of the Langmuir decay process from unmagnetized
to magnetized plasmas by \citet{2013PhRvL.110r5001L} and \citet{2018PhPl...25h2309C} who deduced that
earlier constraints on such process should be removed when including the
magnetization effect.

\begin{figure*}[ht]
 \centering
 \includegraphics[width=0.99\linewidth]{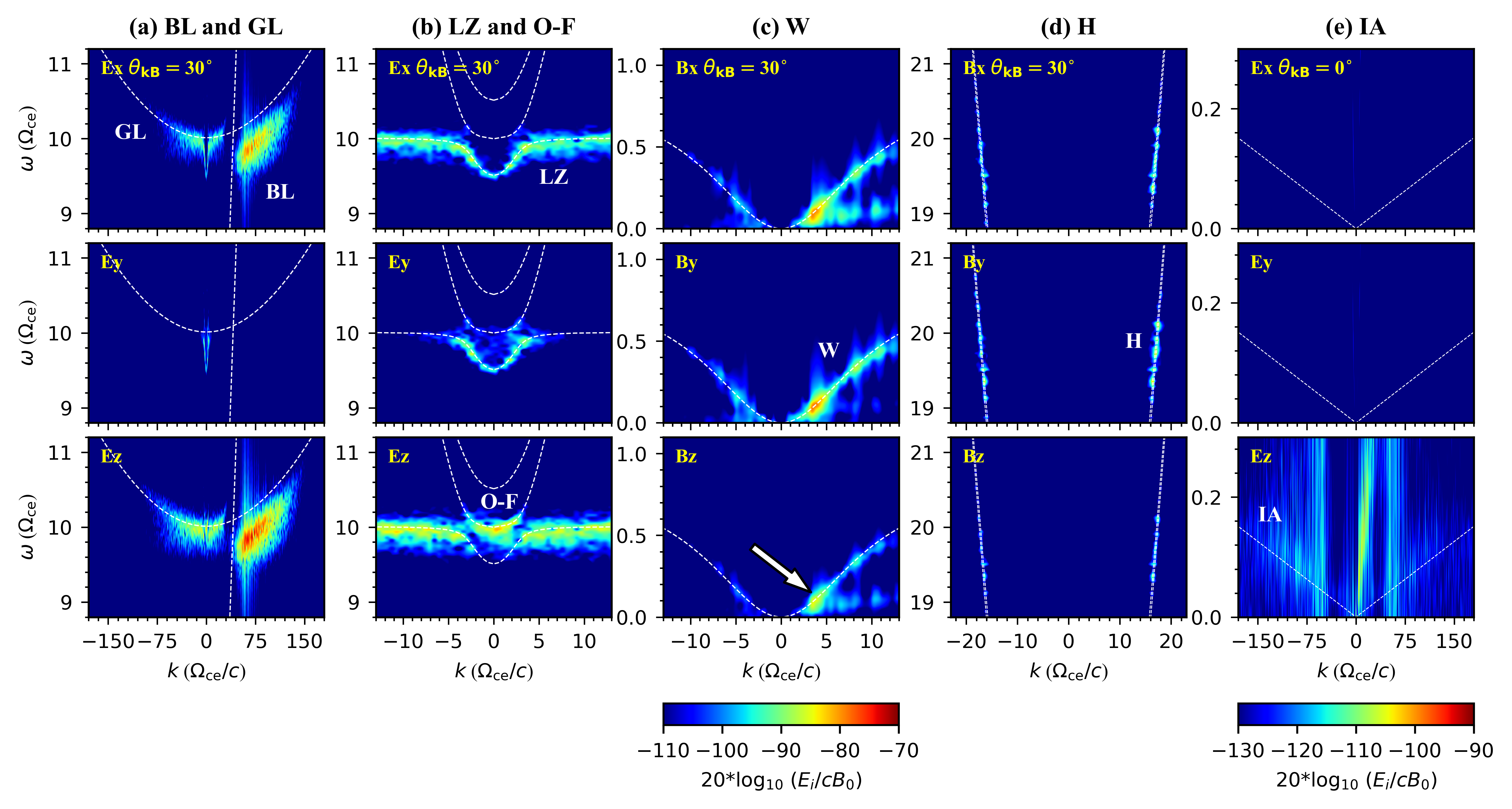}
  \caption{Dispersion diagrams of ($E_x, E_y, E_z$) for the BL (a), GL and O--F modes (b),
those of ($B_x, B_y, B_z$) for the W (c) and H modes (d), and those of ($E_x, E_y, E_z$) for the IA mode. Overplotted lines are corresponding dispersion curves given by the magnetoionic theory. The straight dashed line in panel (a) represents the beam mode. The arrow indicates the spectral location selected to plot the resonance curve (Figure 1) for $\theta = 30^\circ$ (see Figure~\ref{fig1}d) for the W mode. The video begins at $\theta = 0^\circ$ and advances 5$^\circ$ at a time up to $\theta = 90^\circ$. The real-time duration of the video is 5 s.
}\label{fig3}
\end{figure*}

\begin{figure*}[ht]
 \centering
 \includegraphics[width=0.99\linewidth]{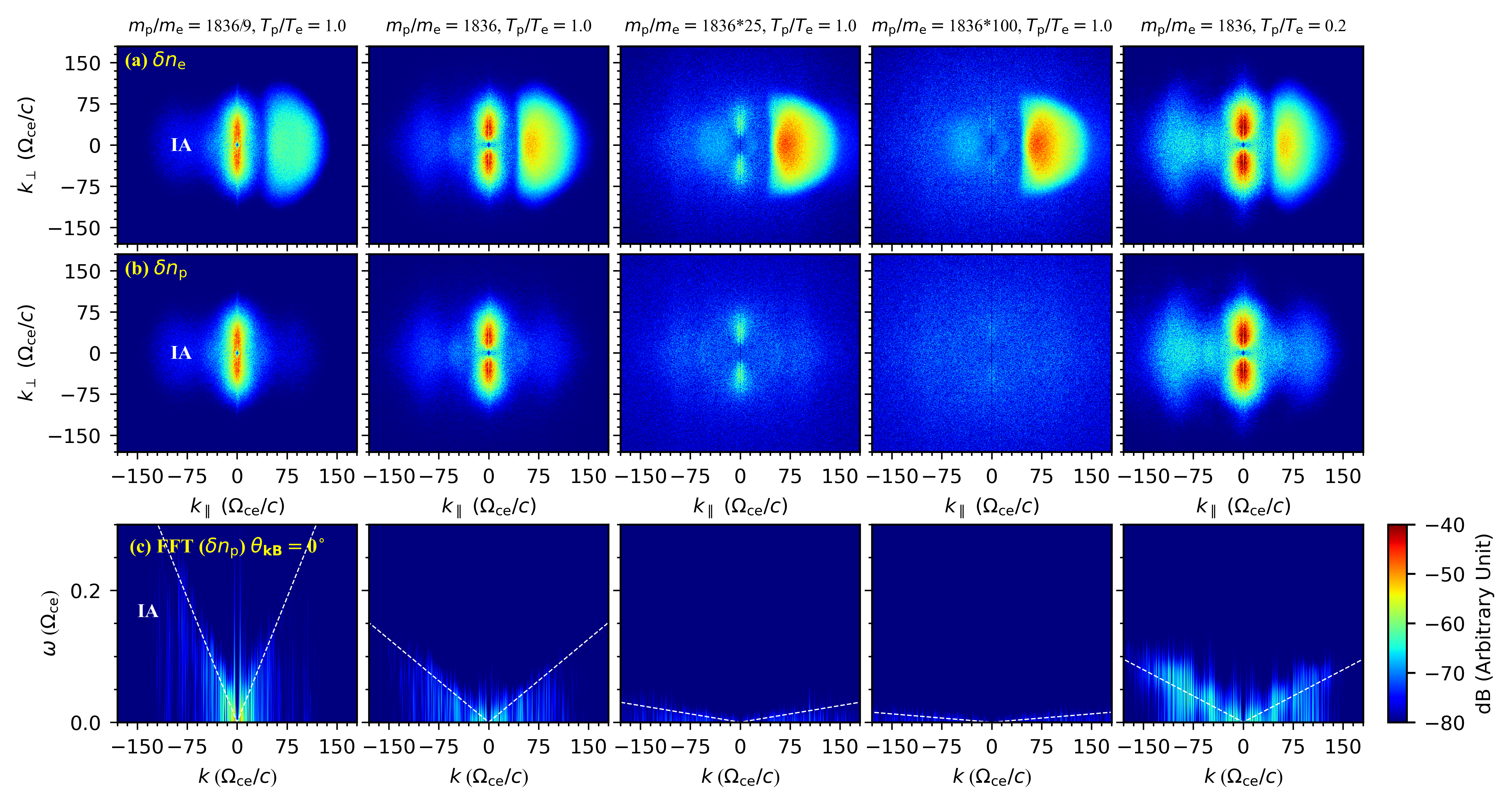}
  \caption{The Fourier spectra in the $\vec k$ space for
density fluctuations of electrons ($\delta n_\mathrm{e}$)
and protons ($\delta n_\mathrm{p}$). The bottom panels
are for the low-frequency regime of the ($\omega, k$) spectra, overplotted dashed lines
are given by the standard IA dispersion relation.}\label{fig4}
\end{figure*}

\begin{figure*}[ht]
 \centering
 \includegraphics[width=0.99\linewidth]{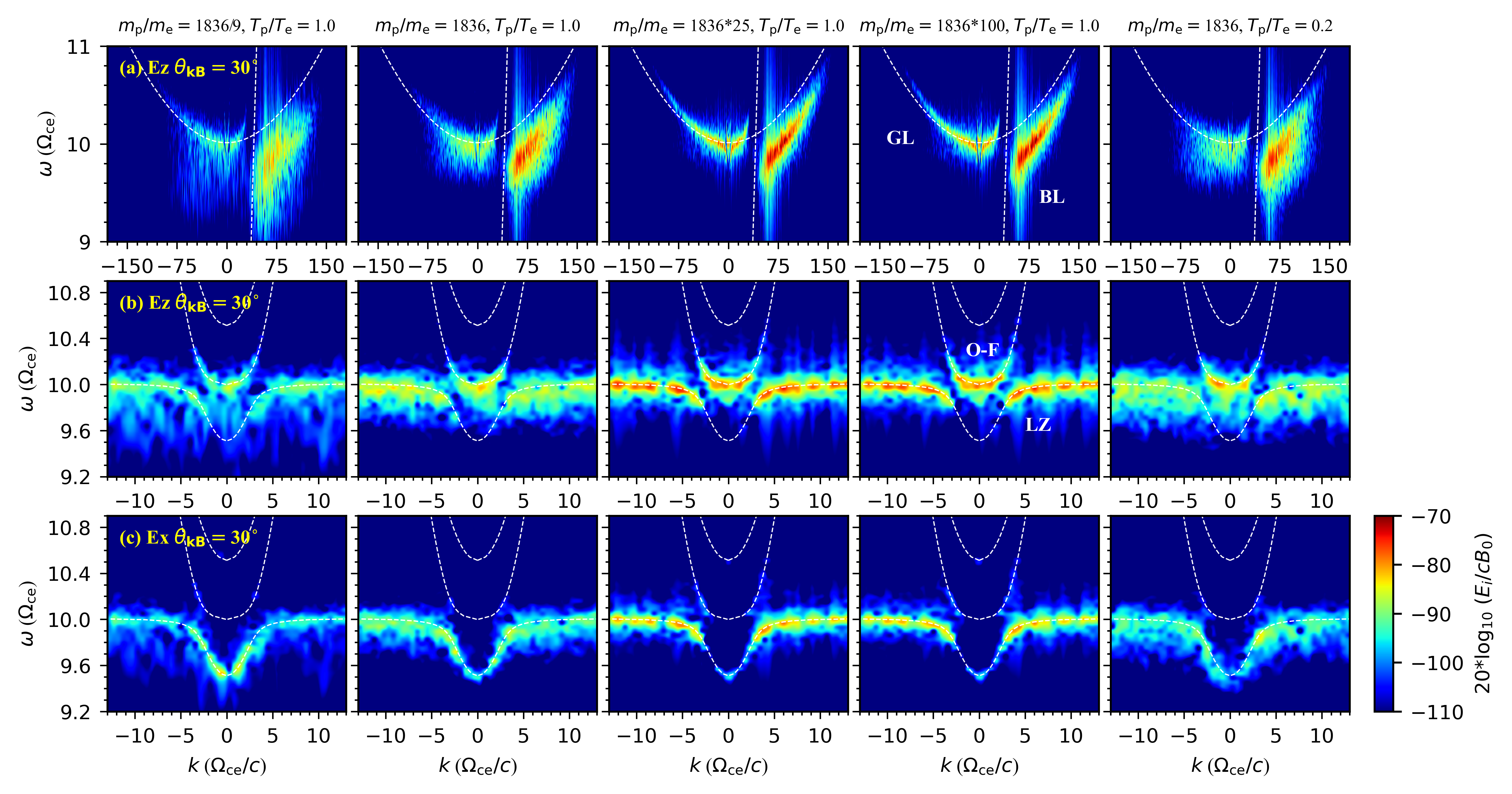}
  \caption{The $\omega-k$ spectra along selected propagating
directions to illustrate the dispersion relation and intensity distribution of the BL, GL, and O--F modes. The video begins at $\theta = 0^\circ$ and advances 5$^\circ$ at a time up to $\theta = 90^\circ$. The real-time duration of the video is 5 s.}\label{fig5}
\end{figure*}

According to the standard PE theory, the H emission is generated by the coalescence
of the primary Langmuir wave and the backward-propagating secondary Langmuir
wave (L + L$\rightarrow$H). In our terms, the process should be expressed as
BL + LT$\rightarrow$H, i.e., coalescence of the beam-Langmuir mode and the backward-propagating thermal
extension of the induced GL mode. Examining Figures~\ref{fig3}--\ref{fig5}, we found that the corresponding matching conditions of $\omega$ and $\vec k$ can be easily satisfied. Thus, we suggest
that the present study supports the standard PE theory of H emission.

The slight enhancement of the O--F mode with increasing IA intensity along quasi-parallel to oblique propagating direction, as shown by comparing
Case R with the ($T_p/T_e = 0.2$) case, may indicate that the coalescence of the BL and the IA mode may
occur and partially contribute to the O--F emission. In addition, an alternative generation mechanism of this mode -- developed recently by \citet{2020ApJ...891L..25N,2021PhPl...28d0701N} and \citet{2021ApJ...909L...5L}
with the same VPIC program for magnetized plasmas yet energized by energetic
electrons with the loss-cone type distribution -- suggests that the O--F emission
is generated via resonant coupling of almost-counter propagating Z and W modes,
i.e., the Z+W$\rightarrow$O--F process. They provided evidence of direct excitation of
the Z mode, and demonstrated the match of the coalescing conditions.
\citet{2021PhPl...28d0701N} verified the occurrence of
such process through dispersion analysis and the wave-pumping PIC simulation.
In the present system, we cannot reject or support
the occurrence of this process since both Z and W modes are excited and
the Z mode can propagate its correlation with the primary BL mode to the O--F mode through the
suspected Z+W $\rightarrow$ O-F process, to be in accordance with the above observations.
In other words, this process may also play a role here. Whether the two mechanisms of
the fundamental PE indeed work together and if yes how much each
contributes remains unaddressed.

\begin{acknowledgments}
This study is supported by NNSFC grants (11790303 (11790300), 11973031, and 11873036). The authors acknowledge Dr. Quanming Lu, Xinliang Gao, and Xiaocan Li for helpful discussion, the anonymous referee for valuable comments, the Beijing Super Cloud Computing Center (BSC-C, URL: http://www.blsc.cn/) for computational resources, and LANL for the open-source VPIC code.
\end{acknowledgments}

\bibliography{chen_zhang2021apjl}{}
\bibliographystyle{aasjournal}

\end{document}